# Nanodiamonds with silicon vacancy defects for non-toxic photostable fluorescent labeling of neural precursor cells

Tobias D. Merson,[1,2,3,*] Stefania Castelletto,[4,*] Igor Aharonovich,[5,*] Alisa Turbic,[3] Trevor J. Kilpatrick,[1,2,3] Ann M. Turnley[3]

[1.] Florey Institute of Neuroscience and Mental Health, & Florey Department of Neuroscience and Mental Health,
Kenneth Myer Building, The University of Melbourne, Parkville, 3010, Victoria, Australia
[2.] Melbourne Neuroscience Institute & [3.] Centre for Neuroscience Research, Department of Anatomy and Neuroscience,
The University of Melbourne, Parkville 3010, Victoria, Australia
[4.] School of Aerospace, Mechanical and Manufacturing Engineering RMIT University, Melbourne, Victoria 3000, Australia
[5.] School of Physics and Advanced Materials, University of Technology Sydney, Ultimo, NSW 2007, Australia

*Corresponding authors: tmerson@unimelb.edu.au, stefania.castelletto@rmit.edu.au, igor.aharonovich@uts.edu.au

*Nanodiamonds (NDs) containing silicon vacancy (SiV) defects were evaluated as a potential biomarker for the labeling and fluorescent imaging of neural precursor cells (NPCs). SiV-containing NDs were synthesized using chemical vapor deposition and silicon ion implantation. Spectrally, SiV-containing NDs exhibited extremely stable fluorescence and narrow bandwidth emission with an excellent signal to noise ratio exceeding that of NDs containing nitrogen-vacancy (NV) centers. NPCs labeled with NDs exhibited normal cell viability and proliferative properties consistent with biocompatibility. We conclude that SiV-containing NDs are a promising biomedical research tool for cellular labeling and optical imaging in stem cell research.*

Fluorescent biomarkers for labeling cellular and molecular targets are emerging as increasingly important tools in biomedical research and medicine. The range of potential applications is diverse, from monitoring drug or tumor localization within the body [1], to assessing the migration of transplanted stem cells used in cell based therapies [2]. A major goal is to develop improved fluorescent labeling reagents that achieve optimal fluorescence intensity without photo-bleaching or blinking. These latter properties are observed with most currently used fluorescent proteins, quantum dots, metallic and dielectric beads and hinders their use for long-term repeated imaging applications [3]. An additional goal is to generate fluorescent biomarkers that can be specifically targeted to distinct cellular or molecular targets via the conjugation of antibodies, growth factors, organic chemicals or drugs.

In these respects, nanodiamonds (NDs) offer several advantages. First, atomic changes in ND structure produces bright optical defects that possess unrivalled photostability, even for the smallest NDs (~5-10 nm) [4,5]. Second, being made of carbon, they are biocompatible, non-toxic and highly amenable to surface functionalization, enabling the conjugation of biomolecules such as DNA and proteins [6-8].

Despite these promising features, further improvement of the optical properties of NDs could enhance their utility as fluorescent biomarkers. The most common optical defect that can be "naturally" incorporated into the NDs is a nitrogen vacancy center (NV), a nitrogen atom close to a vacancy in the diamond lattice [9]. NV emits over a broad range of wavelengths (575 - 800 nm), and has been explored as a potential candidate for atomic resolution magnetic resonance imaging [10]. However despite its promising properties for magnetic sensing, the optical properties of NV centers are not ideal. Specifically, its peak absorption at 532 nm overlaps with wavelengths that excite cellular auto-fluorescence and the defect also possesses a long excited state lifetime of 22 ns that hinders its efficiency. By contrast, NDs containing silicon vacancy (SiV) defects could offer optimal spectral properties for cellular imaging due to their narrow emission line in the near infrared (~FWHM 4 nm centered at 739 nm) and fast excited state lifetime (~1 ns).

In this Letter, we report on the generation of ultra-bright NDs hosting SiV defects and their use as effective biomarkers of primary neural precursor cells (NPCs) isolated from the adult mouse brain. We compare the optical properties and biocompatibility of NDs incorporating NV versus SiV defects and establish the utility of the latter for cellular imaging. The use of SiV-containing NDs to label NPCs is motivated by the need for a bright biomarker that exhibits excellent photo-stability and that is not compromised by cellular auto-fluorescence

The spectral properties of SiV-containing NDs make them ideal candidates for fluorescent biomarkers. The SiV defects consist of a silicon atom in the diamond lattice taking the place of two carbon vacancies [11]. The SiV can be created and observed in ultra-small NDs (sub 5 nm, Ref. 4) [12,13]. Furthermore, since the emission is fully polarized, this in principle allows emission to be tracked by polarization measurements or dipole imaging [14,15]. The combination of these properties makes NDs hosting the SiV defects a prime candidate as an efficient bio-labeling reagent.

To generate fluorescent NDs based on SiV defects, we used two different methods. For Method 1, NDs were grown on a fixed silicon substrate using 4-6 nm seeding NDs to nucleate growth in a microwave plasma chemical vapor deposition (CVD) reactor (Astex) at 900°C, 150 Torr and 950 W followed by annealing for 2 hrs at 1000ºC in forming gas. This achieves a growth rate of approximately 28 nm/min resulting in the production of NDs with an average size of 200 nm after 8 min [16]. Characterization of these NDs by scanning electron microscopy (SEM) revealed uniform crystalline structures (Fig. 1(a)). Due to the fast growth rate the incorporation of silicon from the etched substrate is considerable and strong fluorescence peaking at 738 nm dominated the spectra. In Fig. 1(b) a typical photoluminescence (PL) spectra recorded from the NDs at room temperature using a 532 nm excitation is shown. A sharp peak at 738 nm is the zero phonon line (ZPL) of the SiV. A comparison of the emission spectra of NV-containing NDs is shown.

Fig. 1(c) demonstrates the exceptional photo-stability of fluorescent light emission from excited SiV centers with little to no evidence of fluorescent quenching over a period of 96 minutes of continuous excitation. Very few naturally-occurring NV centers were found in the seed crystal. As a result, grown NDs consist of much higher concentrations of SiV compared to NV.

The second approach to generate SiV-containing NDs (Method 2) was devised to ensure NDs remained in a powder form so that they could be added to cells grown in suspension culture. To achieve this, we utilized an ion accelerator to implant Si into untreated 40 and 80 nm high pressure high temperature (HPHT) NDs ($10^{17}$ Si ions/cm2) at room temperature. After implantation, NDs were annealed for 1 hr in forming gas to create the defects and remove residual graphite. No acid cleaning was performed therefore the fluorescence emission was not optimized.

Compared to the first approach, Method 2 resulted in a much lower yield of SiV defects per ND, but had the distinct advantage that NDs remained in powder form so that they could be added to NPCs grown in suspension culture whereas NDs generated using Method 1 required NPCs to be grown as an adherent monolayer on top of the silicon substrate.

To test the biocompatibility and fluorescent labeling properties of NDs, we established primary cultures of adult mouse NPCs grown as free-floating colonies of cells known as neurospheres. Primary neurospheres were derived from subventricular zone tissue isolated from the brains of four adult C57BL/6 mice that were humanely sacrificed at eight weeks of age (Fig. 2(a)), using procedures approved by the animal ethics committee of the Florey Institute. Single cell suspensions of NPCs were generated by neurosphere dissociation every 5-7 days, using techniques described previously [17].

First we assessed whether labeling NPCs with NDs influenced cell proliferation and/or cell viability. NPCs were plated at a density of 50,000 cells/well in 24-well plates containing proliferation medium and

0, 20, 40 or 100 µg/ml of 35 nm or 100 nm NV-containing NDs (triplicate wells were used for each condition). Fig. 2(b) shows photomicrographs of typical NPC cultures grown either under basal conditions (without NDs) or with 100 µg/ml of 35 nm or 100 nm NDs. NPC-derived neurospheres cultured in the presence of NDs were indistinguishable from NPCs under basal conditions. At concentrations of 100 µg/ml, the uptake of 100 nm NDs into NPCs was evident under light microscopy resulting in a phase dark appearance of cells (Fig. 2(b)).

Plots in Fig. 2(c) reveal that the increase in NPC yield due to cell proliferation over 7 days in culture was unaffected by 100 nm NDs present at up to 40 µg/ml. However, when presented in excess (100 µg/ml), NDs induced a small but significant increase in cell yield relative to the basal condition ($P<0.05$).

We also measured the amount of lactate dehydrogenase (LDH) released into the culture medium after 4 and 7 days culture, to determine whether NDs elicit NPC cytotoxicity. Fifty microliters of supernatant from each well was processed using the CytoTox 96® Non-Radioactive Cytotoxicity Assay (Promega), according the manufacturer's instructions. Plots in Fig. 2(d) revealed that, relative to basal conditions, 100 nm NDs did not influence the level of LDH released into the medium of NPC cultures. Similar results were obtained with 35 nm NDs over the same range of concentrations (data not shown). Collectively, these data demonstrated that NDs had no effect on NPC viability or proliferation, with the exception that 100 nm NDs at 100 µg/ml induced a small increase in NPC yield after 7 days culture.

Next we compared the fluorescence properties of NV and SiV centers for NDs bound or unbound to NPCs using a custom made confocal microscope. NPCs were grown directly on silicon substrates studded with synthesized SiV NDs (generated by Method 1) or by suspension culture in the presence of NV- or SiV-containing NDs (the latter generated by Method 2). For suspension cultures, NPCs were allowed to adhere to laminin-coated glass coverslips overnight then fixed with methanol (-20°C for 10 mins) and mounted onto a glass slide with mowiol mounting medium before imaging. NPCs grown on silicon were fixed in an identical manner. For NV-containing NDs, samples were excited using a 100 mW continuous wave laser operating at 532 nm, with 50 µW power incident on the sample. The laser was focused through the back surface of the coverslip and onto the sample using a 100x infinity corrected oil immersion objective with a numerical aperture of 1.3, and luminescence was collected confocally through a pinhole. A spectrometer with a cooled CCD (Princeton Instruments) was used to characterize the luminescence. Single-photon-sensitive avalanche photodiodes (Perkin-Elmer SPCM-AQR-14) were used to measure the photon count rate for the confocal scan. A 650 nm long pass filter was used to selectively collect the photons from NV center. For imaging SiV in NDs we used a 690 nm CW laser as excitation in a similar confocal system, while a 740 ± 10 nm bandpass filter was used to collect photons from the specific defects. Due to the red excitation wavelength we did not excite NV centers in the NDs.

We initially incubated NPCs with 35 nm and 100 nm NDs containing a high concentration of nitrogen vacancy (NV) centers. Fig. 3 shows epifluorescence imaging of NPCs grown for 7 days in the presence of NV-containing NDs. The broad emission spectra overlapped significantly with intrinsic cellular auto-fluorescence, particularly for NDs of 35 nm. Although the same imaging parameters were sufficient to detect a fluorescent signal from 100 nm NDs, the signal could not be completely separated from that elicited by cellular auto-fluorescence, resulting in low contrast images.

To examine the spectral properties of SiV-containing NDs for cellular labeling, NPCs were cultured for 4 days either as adherent cultures on Silicon substrates studded with 80 nm NDs grown by CVD or as suspension cultures with 40 nm NDs that were implanted with Si. In the first case, the SiV signal was restricted to the Silicon surface indicating limited internalization of substrate-bound NDs by NPCs (data not shown). By contrast, NDs containing SiV defects introduced by Si implantation labeled cells with great efficiency. Fig. 4(a) shows a confocal image collected using an Olympus FV-1000 microscope of NPCs cultured with 40 nm SiV-implanted NDs. The pseudocolored fluorescent signal illustrates the range of fluorescent signal intensities detected among labeled cells. The fluorescent image is overlaid with a bright-field image of the cells collected using Nomarski optics. Fig. 4(b) shows the spectrum recorded from NDs localized intracellularly versus that of NDs found on the outside of the cell. The SiV emission is preserved and the cell does not modify the SiV emission. A comparison with CVD SiV in ND spectra is also shown. The SiV containing NDs offer a superior cell labeling, with very bright photo-stable fluorescence, broad absorption spectra and narrow emission spectra at the near infrared that does not overlap with cellular auto-fluorescence. This analysis revealed the excellent contrast and good signal to noise ratio of SiV-containing NDs, demonstrating the clear advantage of NDs containing SiV compared to NV for cellular imaging.

Our data reveal that the optimal spectral properties of SiV-containing NDs make them excellent candidates for ongoing development as cellular biomarkers. The development of techniques to mass produce NDs with SiV centers is therefore an important consideration. This could be achieved via CVD growth and successive etching of the sacrificial substrate and mechanical milling [13], rather than with ion implantation. In this last case it is as yet unclear if this method could provide a better yield compared to ion implantation. Our study highlights the value in investigating the suitability of NDs containing other color centers for spectral imaging. In particular, NDs with Cr-related centers have similar spectral properties to SiV [18], however mass production of NDs with Cr-related centers is likely to be more problematic than for SiV centers.

In summary, we demonstrate that SiV-containing NDs are excellent biomarkers that have highly favorable attributes compared to NV-containing NDs. Specifically, SiV-containing NDs exhibit bright, photostable emission with a broad absorption spectra and narrow emission spectra in the near infrared range (centered at 738 nm with 4 nm FWHM) which bypasses cellular auto-fluorescence. Further exploration of the utility of SiV-containing NDs for cellular imaging could provide promising avenues for the development of high resolution non-invasive real-time imaging of cell migration in vivo.

*This project was supported by the University of Melbourne interdisciplinary research grant scheme. The Victorian Operational Infrastructure Support Program provided additional support. We thank Dr Brett Johnson for the assistance with ion implantation. Dr. Aharonovich is the recipient of an Australian Research Council Discovery Early Career Research Award (Project No. DE130100592).*

References
1. Y. Urano, Curr. Opin. Chem. Biol. 16, 602 (2012).
2. E. Gu, W.-Y. Chen, J. Gu, P. Burridge, and J. C. Wu, Theranostics 2, 335 (2012).
3. J. Vogelsang, R. Kasper, C. Steinhauer, B. Person, M. Heilemann, M. Sauer, and P. Tinnefeld, Angew. Chem. Int. Ed. Engl. 47, 5465 (2008).
4. I. Vlasov, A. S. Barnard, V. G. Ralchenko, O. I. Lebedev, M. V. Kanzyuba, A. V. Saveliev, V. I. Konov, and E. Goovaerts, Adv. Mater. 21, 808 (2009).
5. I. Vlasov, O. Shenderova, S. Turner, O. I. Lebedev, A. A. Basov, I. Sildos, M. Rähn, A. A. Shiryaev, and G. Van Tendeloo, Small 6, 687 (2010).
6. N. Mohan, C. -S. Chen, H. -H. Hsieh, Y. -C. Wu, and H. -C. Chang, Nano Lett. 10, 3692 (2010).
7. V. N. Mochalin, O. Shenderova, D. Ho, and Y. Gogotsi, Nature Nanotech. 7, 11 (2011).
8. Krueger and D. Lang, Adv. Funct. Mater. 22, 890 (2012).
9. G. Davies and M. F. Hamer, Proc. R. Soc. Lond. A Math. Phys. Sci. 348, 285 (1976).
10. H. J. Mamin, M. Kim, M. H. Sherwood, C. T. Rettner, K. Ohno, D. D. Awschalom, Science 339, 557-560, (2013).
11. J. P. Goss, R. Jones, S. J. Breuer, P. R. Briddon, and S. Öberg, Phys. Rev. Lett. 77, 3041 (1996).
12. E. Neu, D. Steinmetz, J. Riedrich-Möller, S. Gsell, M. Fischer, M. Schreck, and C. Becher, New J. Phys. 13, 025012 (2011).
13. E. Neu, M. Fischer, S. Gsell, M. Schreck, and C. Becher, Phys. Rev. B 84, 205211 (2011).
14. S. Castelletto, I. Aharonovich, B. C. Gibson, B. C. Johnson, and S. Prawer, Phys. Rev. Lett. 105, 217403 (2010).
15. S. Castelletto and A. Boretti, Optics Letters 36, 4224 (2011).

16. Stacey, I. Aharonovich, S. Prawer, and J.E. Butler, Diam. Relat. Mater. 18, 51 (2009).
17. T. D. Merson, M. P. Dixon, C. Collin, R. L. Rietze, P. F. Bartlett, T. Thomas, and A. K. Voss, J. Neurosci. 26, 11359 (2006).
18. I. Aharonovich, S. Castelletto, B.C. Johnson, J.C. McCallum, and S. Prawer, New J. Phys. 13, 045015 (2011).

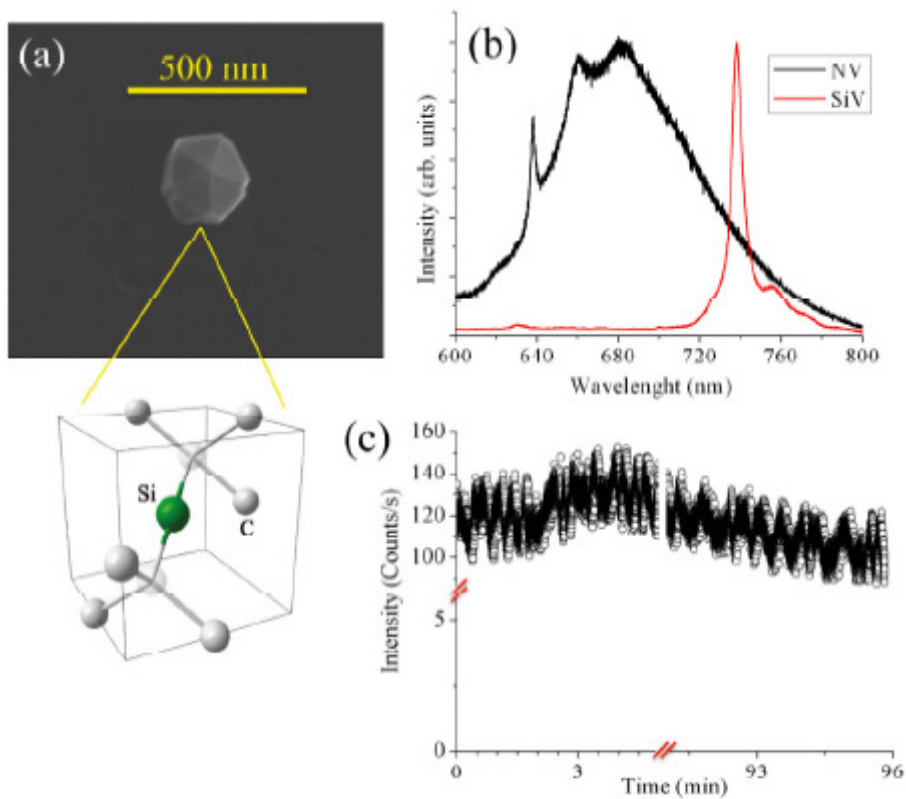

Fig. 1. Structural and spectral properties of SiV-containing NDs. (a) Scanning electron micrographs of a SiV-containing ND synthesized by CVD. Illustration of the Si atom in a di-vacancy interstitial position in the diamond lattice. (b) Typical emission spectra of the SiV observed in CVD NDs compared to the emission of NV centers. The figure clearly illustrates the advantage of the SiV for bio-labeling due to its narrowband fluorescence in the near infrared range. (c) Plot of photo-stability of SiV emitted fluorescence following continuous excitation (532 nm) for over 1.5 hours. Fluctuations in intensity were merely due to laser instability that occurred during only these specific measurements.

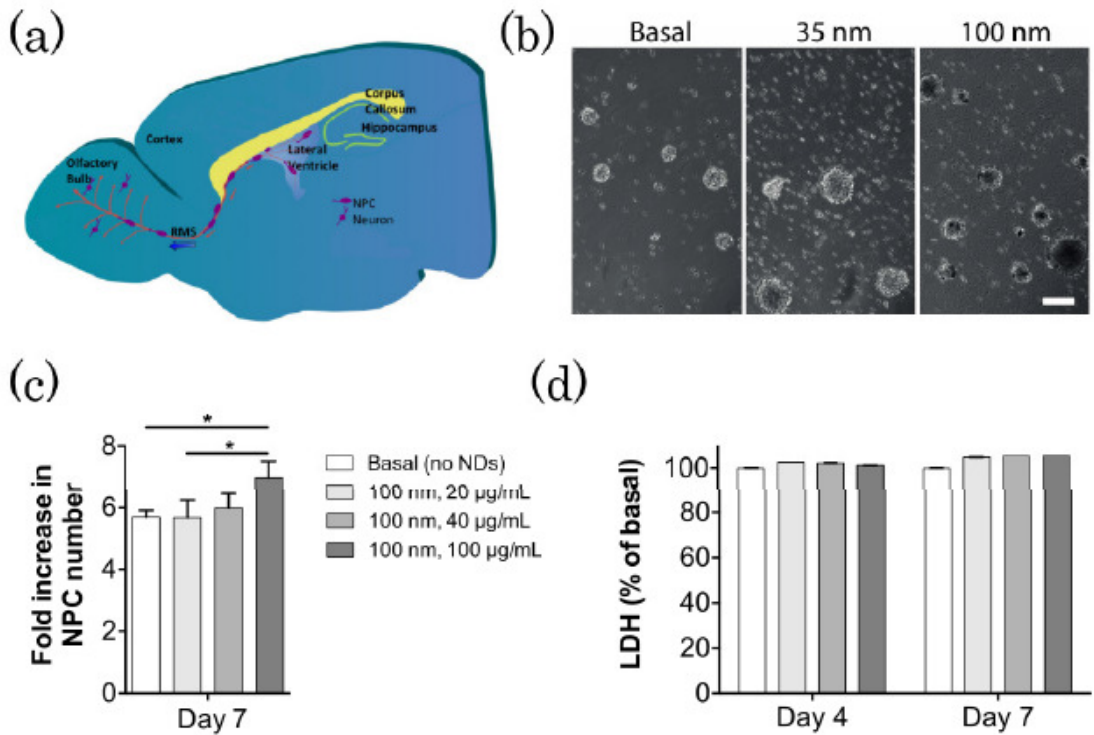

Fig. 2. NPCs isolated from the adult mouse brain exhibit high biocompatibility with NDs. (a) Schematic view of the adult mouse brain in sagittal axis. Primary NPCs were isolated from the subventricular zone of the lateral ventricles. NPCs contribute to neurogenesis by migrating to the olfactory bulb via the rostral migratory stream (RMS). (b) Morphology of NPC-derived neurosphere cultures assessed 7 days after culture with NDs added at time of passaging. Scale bar: 100 µm. (c) NPCs labeled with 20 or 40 µg/ml NDs exhibited normal increases in cell yield. NDs at 100 µg/ml caused a small but significant increase in cell yield relative to basal ($P<0.05$, one-way ANOVA, Tukey post-hoc analysis). (d) Assessment of cell death of NPCs cultured in the presence of NDs using the LDH cytotoxicity assay. After 4 and 7 days of culture NPCs cultured with NDs had LDH levels that were equivalent to those cultured under basal conditions without NDs indicating that NDs are biocompatible and non-toxic.

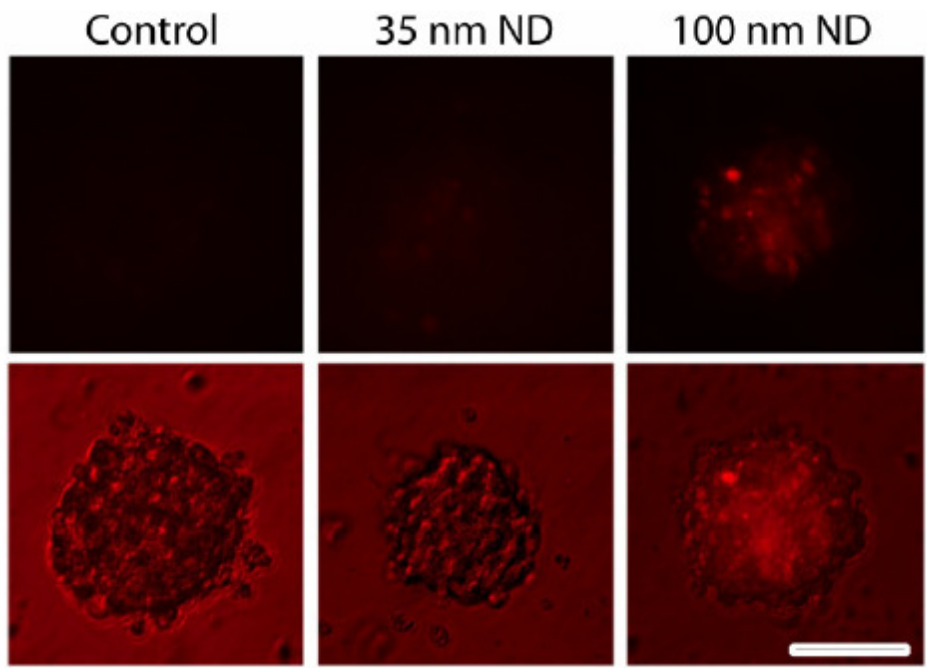

Fig. 3. Epifluorescence image of NPCs grown with enhanced NV centres containing NDs (upper images) and dark field images (bottom) of the same sample area. The NDs incubated with the cells were 35 and 100 nm respectively. The 35 nm NDs show a limited emission compared to the 100 nm ND. Scale bar: 50 µm.

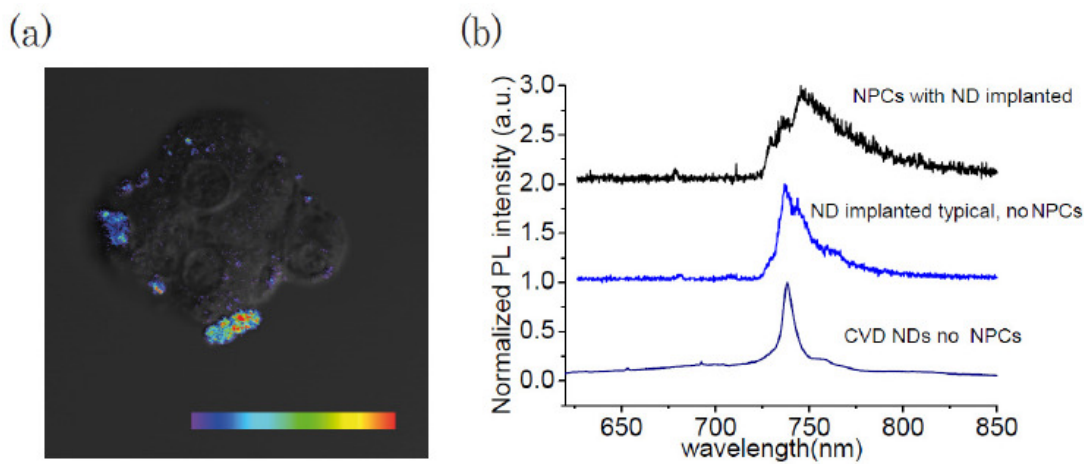

Fig. 4. (a) A confocal image of NPCs grown on NDs merged with a bright field image of the neurosphere in grayscale collected by Nomarski optics revealing the exceptional contrast of SiV-containing NDs over cell background. (b) PL of SiV in CVD NDs and NDs implanted with Si in cell culture and without cells. The broader emission of the SiV in the cultured cell sample were attributed to the use of a thick cover glass for which the objective was not fully corrected, inducing a larger focal volume collection compared to the case where only NDs were imaged.